\documentclass[10pt,final,journal,a4paper,oneside,twocolumn]{IEEEtran}
\usepackage{amssymb}
\usepackage[tbtags,sumlimits,nointlimits,reqno]{amsmath}

\usepackage[utf8]{inputenc}
\usepackage{bbm}
\usepackage{float}
\usepackage{srcltx}
\usepackage{color}

\usepackage{color}
\usepackage{mathdots}
\usepackage[export]{adjustbox}
\usepackage{amsmath}
 \usepackage{dblfloatfix}

\usepackage{color}
\usepackage{float}
\usepackage{mathtools, cuted}
\usepackage[makeroom]{cancel}
\usepackage{relsize}
\usepackage{bbm}
\usepackage{ulem}
\usepackage{cite}
\usepackage{booktabs}
\usepackage{multirow}

\makeatletter
\let\MYcaption\@makecaption
\makeatother

\usepackage[font=footnotesize]{subcaption}

\makeatletter
\let\@makecaption\MYcaption
\makeatother

\definecolor{amber}{rgb}{0.8, 0.33, 0.0}

\newcommand{\blue}[1]{\textcolor{black}{#1}}

\definecolor{myGreen}{rgb}{0.1137, 0.5333, 0.2823}
\definecolor{myRed}{rgb}{0.8196,0.0588,0.0705}
\newcommand{\jn}[1]{\textcolor{black}{#1}}

\newcommand{\parallelTwo}[2]{\left.#1\,\middle|\!\middle|\,#2\right.}
\newcommand{\parallelThree}[3]{\left.#1\,\middle|\!\middle|\,#2\,\middle|\!\middle|\, #3\right.}

\usepackage{flushend}

\begin{document}

\normalem

\title{Millimeter-Wave Huygens' Transmit-Arrays based on Coupled Metallic Resonators}

\author{%
        Soichi Sakurai, Jo\~{a}o G. N. Rahmeier, Takashi Tomura, \\Jiro Hirokawa and Shulabh Gupta
\thanks{This work was supported in part by Support Center for Advanced Telecommunications Technology Research Foundation (SCAT) and Mizuho Foundation for the Promotion of Sciences.}
\thanks{Jo\~{a}o G. N. Rahmeier and S.~Gupta is with the Department of Electronics, Carleton University, Ottawa, Ontario, Canada. (email: shulabh.gupta@carleton.ca)}
\thanks{S. Sakurai, T. Tomura and J. Hirokawa are with the Department of Electrical and Electronic Engineering at Tokyo Institute of Technology, Tokyo, Japan (email: tomura@ee.e.titech.ac.jp)}
}

\markboth{MANUSCRIPT DRAFT}
{Shell \MakeLowercase{\textit{et al.}}: Bare Demo of IEEEtran.cls for Journals}

\maketitle
\begin{abstract}
A novel Huygens' transmit-array is proposed based on \jn{a} coupled-resonator approach where two identical elliptical metallic patches with an elliptical hole separated by a dielectric substrate has been used to demonstrate millimeter-wave (mm-Wave) beam-forming for linear polarization. The proposed structure is simple and compatible with standard Printed Circuit Board (PCB) processes and utilizes a single dielectric substrate only. It is shown that by engineering the geometrical dimensions of the resonator, its electric (even-mode) and magnetic (odd-mode) resonances are excited in a balanced manner to achieve zero back-scattering in a large bandwidth. This operation principle of the proposed Huygens' cell is explained in details using both an insightful equivalent circuit model, as well as full-wave \jn{eigenmode} analysis. Next, the proposed Huygens' cell is placed on top of a high gain 2D slot-array antenna, \jn{in its near-field,} to engineer its aperture field distribution, where the resulting Huygens' transmit-array acts as a broadband phase plate. Several transmit-array prototypes designed around the 60 GHz frequency band are demonstrated and experimentally characterized in both their near and far-fields, to achieve difference pattern generation, beam expansion and beam steering as application examples, in addition to a uniform surface demonstrating its low-loss performance. Further discussions related to the unit cell size vs frequency bandwidth trade-offs and future extension to handling circular polarization are finally provided. 
\end{abstract}



\section{Introduction}

There recently has been a resurgence in the area of Electromagnetic wave propagation engineering and control using engineered surfaces with extensive amplitude, phase and polarization control. Using \jn{a versatile and powerful platform} of electromagnetic metasurfaces, wide variety of novel applications have emerged, which has greatly enhanced the wave manipulation capabilities of traditional transmit- and reflect-arrays and Frequency Selective Surfaces (FSSs). These metasurfaces are two dimensional structures \jn{that consist} of sub-wavelength resonators, similar to transmit-arrays and FSSs, which due to their engineerable geometrical features provide superior desired macroscopic responses, which otherwise are not possible \cite{Holloway_Metasurface, Tretyakov_MS, Tretyakov_Huygens, Yu_PhaseControl, Glybovski_PhaseControl}. Wide variety of wave transforming applications has been proposed such as cloaking \cite{Gharghi_Cloak}, perfect absorption \cite{Moitra_Reflection, BBParticlesTratyakov}, polarization control \cite{Shi_PolarizationControl} \cite{Grbic_Metasurfaces} and holograms \cite{MetaHolo}, to name a few. 

Approaching from the application perspective, with recent and rapid developments in the next generation wireless technologies such as 5G, there has been a greater need for advanced wave manipulation and transformation \cite{Tech_5G_Challenge, 5G_mmWave}. There has been an active push to develop novel technologies in the millimeter-wave frequencies, such as 60 GHz in IEEE 801.11ad, due to large spectrum availability \cite{60WaveTech, mmWaveTech}. This is reflected in renewed interest in the development of new antenna systems and phased array antennas, for instance, for high-speed data communication system. An example of it is a 60-GHz-band Gigabit Access Transponder Equipment (GATE) system, which uses the near-field region to provide stable field intensity to enable ultrahigh speed communication link \cite{GATE}. 

GATE system has been demonstrated using large-scale slot array antennas based on diffusion bonding processes, achieving high efficiency antennas, and suitable for communication ranges in the order of tens of meters \cite{Tomura_Dif_Antenna1, Tomura_Corporate, DiffusionBond}. However, due to complexity of these antennas in terms of both design sophistication and \jn{sensitive fabrication processes}, they have been restricted to uniform apertures with optimal broadside radiation beams in a large bandwidth. \jn{Nevertheless, there are requirements} to engineer the radiation characteristics of these antenna to achieve diversity of beam \jn{coverage} in the near-field region of the antenna apertures. Examples include, beam focussing, beam tilting, non-uniform coverage, for instance, which requires different aperture field distributions to generate corresponding radiation patterns. Therefore, engineering the aperture fields of the large-scale slot array antennas, without modifying the antennas themselves, represents an ideal situation to exploit the wave transformation capabilities of electromagnetic metasurfaces and related transmit-array structures.

\begin{figure*}[htbp]
\begin{center}
\includegraphics[width=2\columnwidth]{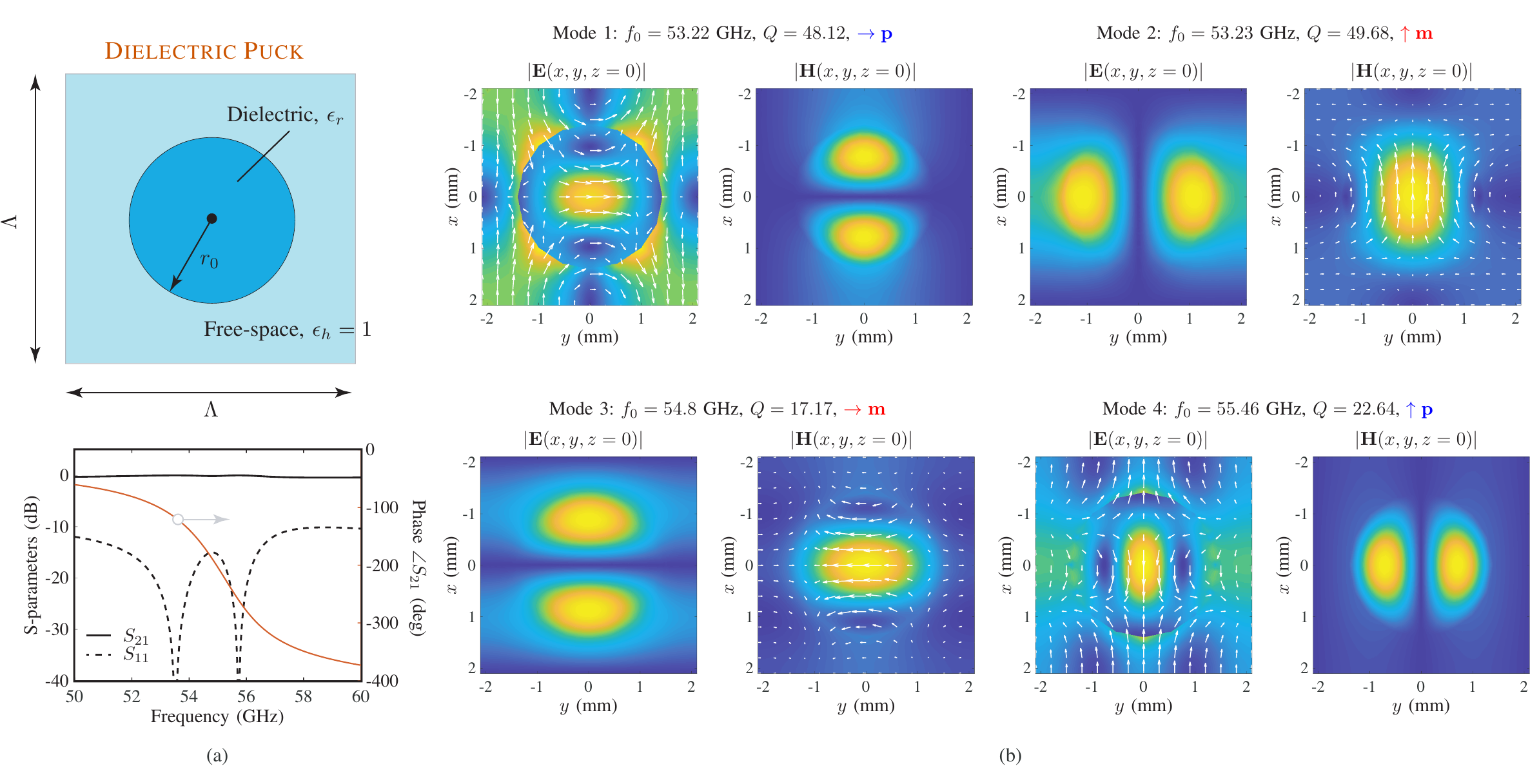}
 \caption{Conventional all-dielectric Huygens' source and its typical electromagnetic response. a) Unit cell consisting of a high permittivity dielectric puck embedded inside free space, and its typical transmission and reflection response (FEM-HFSS, driven simulation). b) Corresponding eigenmodes of the puck (FEM-HFSS, eigenmode simulation). Simulation parameters are: $\epsilon_r = 6.5$, $r_0=1.4$~mm, $h=1.27$~mm, $\Lambda = 4.2$~mm.}\label{Fig:DieelctricHuygensCell}
\end{center}
\end{figure*}

This objective of engineering the antenna radiation can be visualized as a spatial cascade of the slot-array antenna structure and a Transmit-array, which is engineered as a phase plate with zero back-reflection to the antennas \cite{Elef_LWA_MS}\cite{Gupta_mm_MS}. The spatial variation of the phase is then used to engineer the field distribution, which in turn affects the radiation coverage characteristics of the resulting compound structure. Huygens' metasurfaces are thus ideal candidates to perform this function due to their unique zero back-scattering properties in a large bandwidth, due to an optimal interaction of its electric and magnetic dipolar moments \cite{Kerker_Scattering}. A lot of work has been done in developing high efficiency Huygens' surfaces, particularly in optics \cite{Howes_AD,Nuaimi_AD,Decker_AD, Saman_Huygen}, while few works have recently emerged in the microwave and mm-Wave frequency ranges. Huygens' surfaces can be realized either using all-dielectric \cite{Achouri,Howes_AD,Nuaimi_AD,Decker_AD,Saman_Huygen} or metallic resonators \cite{Shengli_MS,Xue_Huygens,Kim_Huygens}.

\jn{In the context of the sought} after GATE application, few Huygens' metasurfaces in the mm-Wave have been reported in the literature, particularly in the 60 GHz bands \cite{Xue_Huygens, GrbicmmWave, mmWaveHologram}. While dielectric resonator based Huygens' surfaces and dielectric surfaces require non-standard processes such as dielectric machining \cite{Sun,PhaseshiftingSurface, Emara}, typical metallic implementations require multi-layer printed circuit board (PCB) processes \cite{GrbicmmWave, mmWaveHologram}. \jn{Recently, double layer PCB based metasurfaces were engineered to emulate a Huygens' response with more than $2\pi$ phase coverage, and a good performance around 30 GHz frequency band, albeit at an expense of higher reflection losses \cite{Xue_Huygens, SingleDielectricHUygen}. Also, they are limited to linear polarization only, and have been operated in the far-field of the horn antenna source. These characteristics are important in the context of the GATE system, for instance, because it utilizes circular polarization and needs to be compact. Therefore, it requires any Huygens' transmit-array to be scalable to handle any polarization and to be placed closer to the large-scale antenna array structure for compactness and near-field operation.}

In this work, a novel Huygens' cell is proposed as the basic building block of the mm-wave transmit-array, operating in the near-field of a linearly polarized slot-array antenna around 60 GHz. The proposed Huygens' cell is based on two coupled resonators, which emulates electric and magnetic dipole resonances to provide broadband matching to free-space, along with engineerable transmission phase, as required in the GATE system. The fundamental working principle of the proposed structure is explained in details using \jn{eigenmode analysis}, along with an intuitive and insightful equivalent circuit model. It is next placed in the near-field of a 2D slot array antenna, and its wave transformation performance is experimentally demonstrated using several examples of beam-tilting, beam expansion and difference pattern generation. Consequently, there are two major contributions of this work: a novel PCB based Huygens' cell along with its equivalent circuit model, and its application as a transmit-array on top of a 2D slot array antenna in its near-field to engineer \jn{its radiation pattern characteristics} around 60 GHz.

The paper is structured as follows. Sec.~II shows the unit cell configuration of the proposed Huygens' cell, along with its detailed eigenmode analysis, in comparison to an ideal planar Huygens' cell based on dielectric resonators. It further presents an equivalent circuit model of the structure, which is shown to correctly reproduce the electromagnetic response of the physical cell. Sec.~III presents the principle of Huygens' transmit-arrays on top of a uniform 2D slot array antennas and the corresponding design procedure for engineering the phase of the Huygens' cell \jn{in such conditions}. It further outlines the key matching and transmission phase characteristics of the structure. Sec.~IV shows various prototypes and presents their measured near- and far-field characteristics for a variety of wave transformations. Some discussions are provided in Sec. V, showing the inherent trade-off between matching bandwidth and unit cell period, and possible extension to handle circularly polarized waves. Conclusions are finally provided in Sec. V, with few additional remarks about the proposed Huygens' structure.

\section{Proposed Huygens' Structure}

\subsection{Huygens' Unit Cell}

A Huygens' surface is composed of a 2D array of Huygens' cells (say in $x-y$ plane), which consists of co-located electric and magnetic dipoles which are orthogonally polarized in space. \jn{Their mutual interaction} results in a destructive interference along one direction (along $-z$), while the two contributions add up in phase in the $+z$ direction \cite{Kerker_Scattering}. Such an array when excited with a uniform plane-wave, exhibits zero back-scattering typically in a large bandwidth, along with finite amount of transmission phase shift that depends on the exact geometrical characteristics of the Huygens' structure. \jn{Therefore, it acts as an ideal phase plate for phase engineering applications.} 

A very simple realization of a Huygens' configuration in a planar form is using a \emph{dielectric resonator} as shown in Fig.~\ref{Fig:DieelctricHuygensCell}(a). It simply consists of a circularly-cylindrical high permittivity dielectric resonator surrounded by a low-permittivity host medium (e.g. air). \jn{The graphic at the bottom of Fig.~\ref{Fig:DieelctricHuygensCell}(a) show its typical reflection and magnitude characteristics, along with transmission phase, when excited with a uniform plane-wave.} Here the structure has been properly optimized to superimpose the electric and magnetic resonances\jn{, forming the Huygens' configuration.} Floquet boundaries are used to emulate a perfectly periodic array along $x$ and $y$ dimensions. As can be seen, it provides a good matching in a large bandwidth, with particularly two distinct reflection nulls. A perfect Huygens' configuration is formed at these two frequencies with balanced electric and magnetic dipolar responses providing a near $2\pi$ phase coverage \jn{across the frequency range} \cite{Grbic_Metasurfaces}.

To further understand the operation of a Huygens' cell, let us consider its eigenmode field distribution. Fig.~\ref{Fig:DieelctricHuygensCell}(b) shows the eigenmode analysis of the dielectric resonator cell using FEM-HFSS. The simulation model consists of perfect magnetic conductor (PMC) boundary along the $x-$direction (to support $x-$polarized excitation), and free-space impedance boundaries along $\pm z$ axis. \jn{In the frequency range of interest (based on the driven response), near the two reflection nulls, four distinct modes are identified. Around each reflection null,} a mode pair is found with similar Q-factors, where each of the field distributions corresponds to either an electric (even-mode) or a magnetic-type (odd-mode) mode. \blue{For example, mode 1 exhibits a $y-$polarized electric field in the center region of the cell, similar to a horizontal electric dipole, while becoming tangential to the left and right side walls to satisfy the PEC boundary condition (accounting for intercellular coupling). In the plane of the puck, the electric field is symmetric about $x=0$. The magnetic field on the other hand circulates in the $x-z$ plane about $x=0$, and is dominantly a $z-$oriented field (shown as a magnitude plot only since the normal vector has poor visibility). This configuration thus forms an electric dipole denoted as a $y-$polarized $\mathbf{p}$. This is an even mode since the E-fields are symmetric about the $x-y$ plane. Mode 2, on the other hand, shows strong H-field components polarized along the $x-$axis, similar to an $x-$polarized magnetic dipole, symmetric about the $y=0$ axis. The corresponding E-fields circulates around $y=0$, and is $z$-polarized. This forms the $x-$polarized magnetic dipole (odd-mode) response of the puck, denoted as $\mathbf{m}$. This is an odd mode since the E-fields are asymmetric about the $x-y$ plane. The puck thus supports two collocated orthogonal dipole moments $\mathbf{p}_y$ and $\mathbf{m}_x$, which following superposition of their corresponding even and odd mode field configurations, produce destructive interference along one direction of $z-$axis, and constructive interference along the other. This is the Huygens' field configuration at the first reflection null of the driven frequency response. Similar conclusions can be made for the other pair of modes at the second reflection null.}

We should note that the dielectric Huygens' structures (reported in the literature) typically exhibit large unit cell periods, $\Lambda$, and are widely used in the context of metasurfaces. Before proceeding further, we would like to clarify this terminology. Due to their common feature of wave transformation functionalities, the conventional transmit-/reflect-arrays are closely related to the general paradigm of electromagnetic metasurfaces. However, an important difference lies in their respective unit cell periodicities. While metasurface unit cells are ideally deeply sub-wavelength to allow homogenization description, transmit-arrays typically exhibit unit cell sizes greater than $\lambda/4$ \cite{MS_rev_Transmit}. Nevertheless, several works have continued to refer Huygens' cells with large unit cell periodicities as metamaterial/metasurface structures, which has blurred the semantics in the relevant literature \cite{Saman_Huygen}. \jn{Thenceforward, we will follow the unit cell periodicity criterion and refer to these unit cells with large periodicities as \emph{Transmit-arrays}.}

\begin{figure*}[htbp!]
	\centering
\includegraphics[width=2\columnwidth]{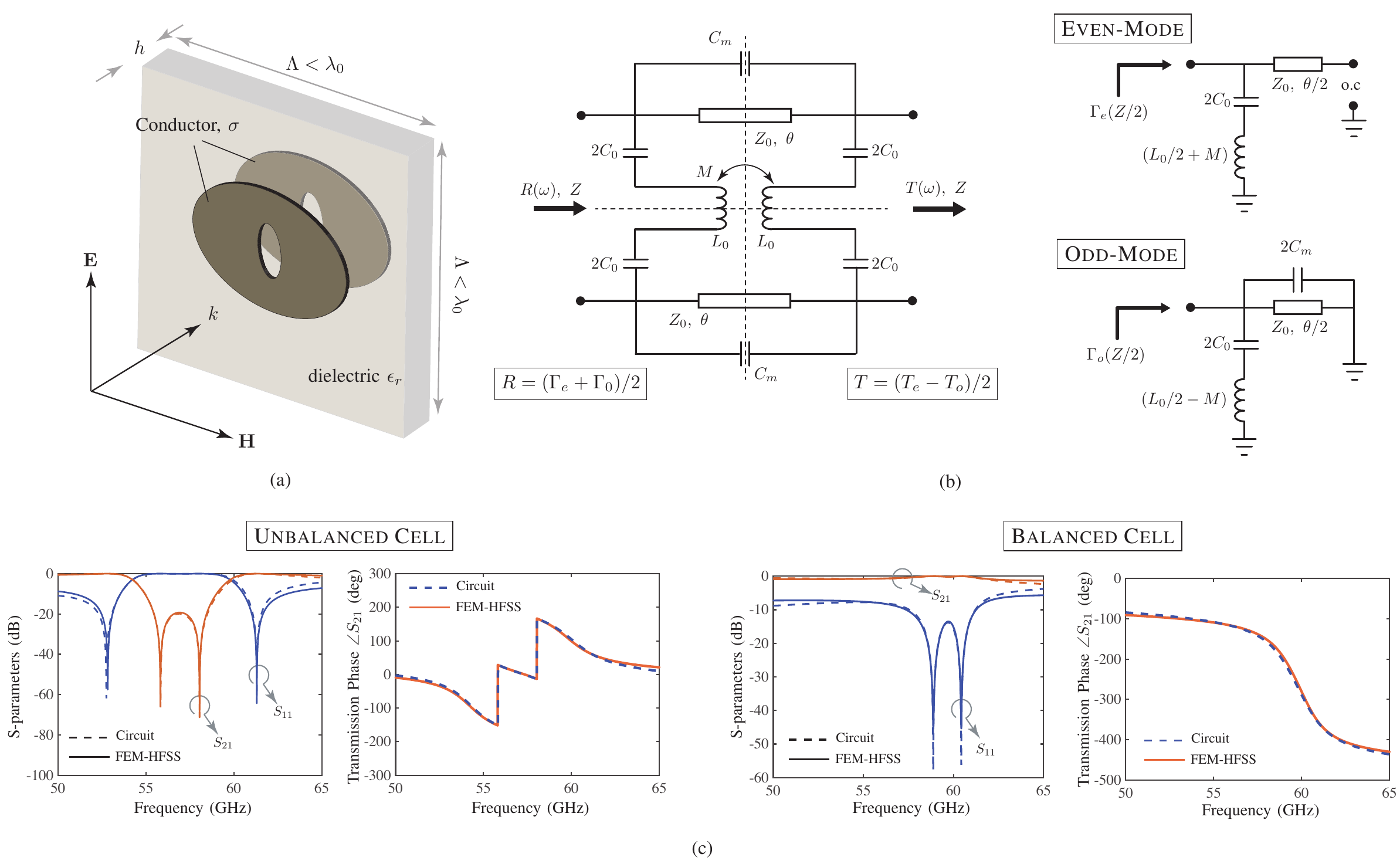}
	\caption{Proposed Huygens' cell as a unit element for constructing a broadband matched transmit-array. (a) Unit cell configuration consisting of two coupled resonators separated by a thin low-permittivity dielectric. (b) Equivalent circuit model and its even-odd mode decomposed circuit, for an assumed lossless cell. (c) S-parameter response of a typical proposed Huygens' cell and its comparison with the circuit model with extracted element values. Simulations parameters are: Unbalanced case -  $R_\text{out} = 0.9$~mm, $R_\text{in} = 0.3$~mm with circuit parameters $C_0= 0.805$~fF, $L_0= 9.677$~nH, $C_m= 6.796$~fF, $M = 0.1222L_0$; Balanced case - $R_\text{out} = 0.82$~mm, $R_\text{in} = 0.25$~mm with $C_0= 0.683$~fF, $L_0= 10.35$~nH, $C_m= 5.938$~fF, $M = 0.0364L_0$. Transmission line impedance $Z_0 = \eta_0/\sqrt{\epsilon_r}$, and $\theta = 55.83^\circ$ at 60 GHz. For the FEM-HFSS simulation, the structure is a ring of inner and outer radii of 0.25~mm and 0.82~mm, respectively for balanced and 0.3~mm and 0.9~mm for unbalanced design. Unit cell size $\Lambda = 4.2$~mm, $\epsilon_r =  2.33$ and $h=0.508$~mm are same for both cases.}
	\label{Fig:CoupResHuyCell}
\end{figure*}

\subsection{Coupled Resonator Huygens' Cell}

\jn{Although the idea of an ideal dielectric resonator Huygens' cell is simple, it is challenging to implement it.} While some dielectric structures have been proposed at the microwave frequencies such as in \cite{Achouri}, based on dielectric bridged interconnections, they are based on non-standard fabrication processes and are particularly difficult to realize at mm-Wave frequencies. 

\begin{figure*}[btp]
\begin{center}
\includegraphics[width=2\columnwidth]{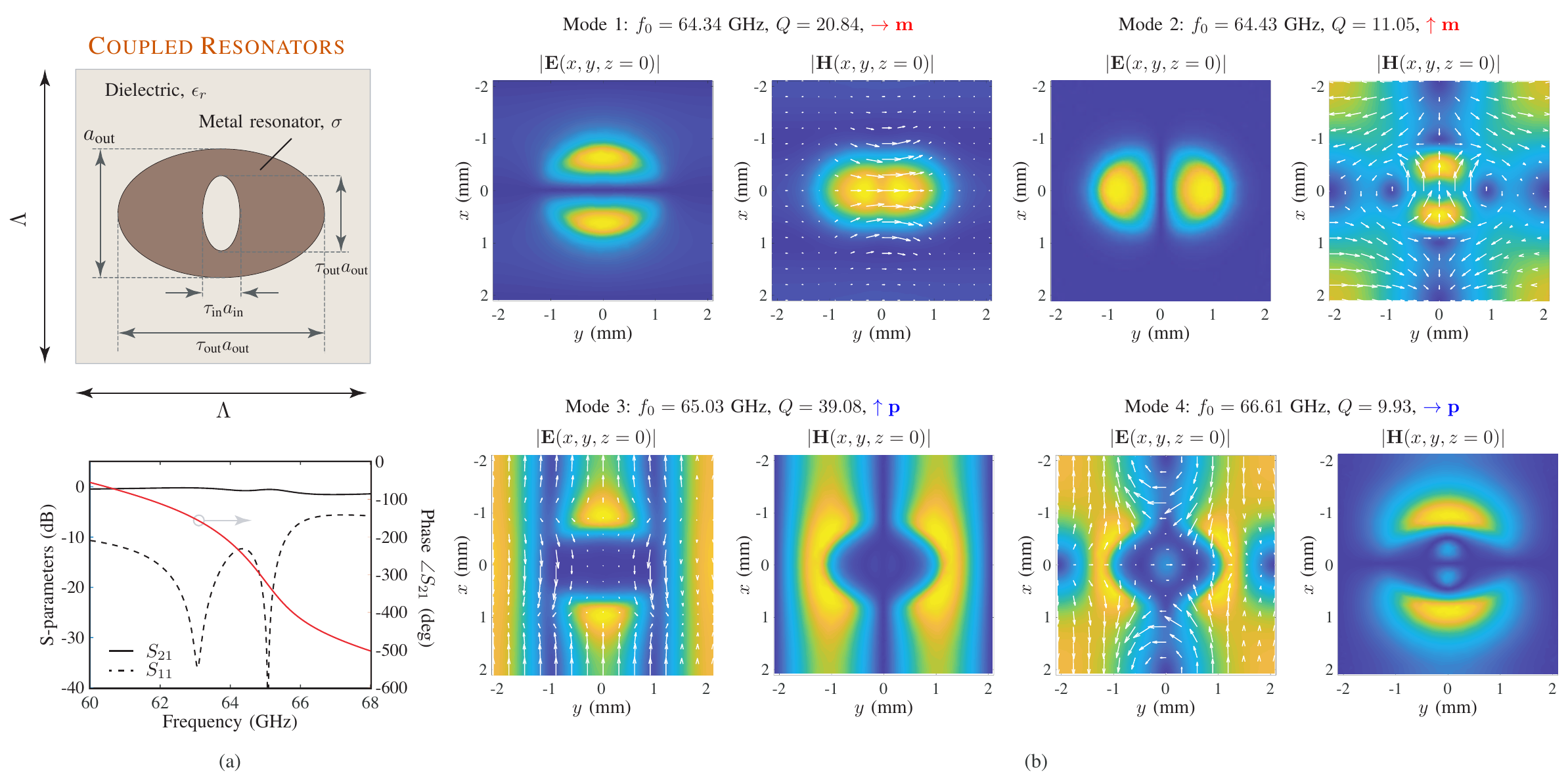}\caption{Proposed Huygens' source based on coupled resonators and its typical electromagnetic response. a) Unit cell consisting of a pair of metallic patches across a dielectric slab, and its typical transmission and reflection response (FEM-HFSS, driven simulation). b) Corresponding eigenmodes of the coupled resonators (FEM-HFSS, eigenmode simulation). Simulation parameters are: $\epsilon_r = 2$, $\tau_\text{out} = 1.5,~a_\text{out} = 775~\mu$m, $\tau_\text{in} = 0.5,~a_\text{in} = 400~\mu$m puck thickness $h=0.508$~mm, $\Lambda = 4.2$~mm and $h=0.508$~mm.}\label{Fig:EigenHuygensCell}
\end{center}
\end{figure*}

The core of a Huygens' cell is the interaction of even and odd modes that provides perfect field cancellation along one direction, i.e. zero back-scattering. Another simple, yet practical, approach to realize such configuration is a \emph{coupled resonator}, as proposed in Fig.~\ref{Fig:CoupResHuyCell}(a). It consists of a metallic patch, separated by a thin host dielectric. \blue{Here, an elliptical shaped patch resonator is proposed to vary the unit cell response, controlled by the inner diameter of the hole $a_\text{in}$, outer diameter $a_\text{out}$ with two inner and outer ellipticities $\tau_\text{in}$ and $\tau_\text{out}$, respectively, as illustrated in Fig.~\ref{Fig:EigenHuygensCell}(a). This provides sufficient degrees of freedom for controlling its scattering response for an $x-$polarized field excitation. While, the outer ellipticity and the diameter primarily primarily controls the location of the reflection nulls in frequency, the inner hole dimensions provide fine tuning to achieve broadband matching.} The unit cell period $\Lambda < \lambda_0$ to ensure sub-wavelength operation of the cell. In addition to the self-coupling of individual metallic patch resonators, additional resonances are generated due to tight electromagnetic coupling through the thin substrate. As will be shown later using the eigenmode field analysis, this configuration is capable of providing even and odd mode field distributions, similar to that in the dielectric resonators, as desired to achieve a Huygens' response.

A simple way to illustrate the proposed cell is using an equivalent circuit model, as shown in Fig.~\ref{Fig:CoupResHuyCell}(b) for a lossless cell, for simplicity. An incoming plane-wave induces an electric current on the metallic patches, which is represented by an inductance, $L_0$. A series capacitance also exists between two neighboring patches in the same plane of the surface, represented by a capacitance $C_0$. This series resonator $\{L_0,~C_0\}$ thus represents the metallic resonator and the equivalent circuit consists of two of these resonators separated by a transmission line of impedance $Z_0$ and electrical length $\theta$, representing the dielectric substrate. The two resonators are further electromagnetically coupled - both inductively and capacitively. A coupling capacitance $C_m$ exists between the two metallic patches along with a mutual inductance of value $M$. \jn{The circuit elements are further split to emphasize both transverse and longitudinal symmetries of the structure.}

The two symmetries, and in particular the longitudinal one, allows an even-odd mode decomposition of the structure. The equivalent circuit model of the respective even and odd modes are also shown in Fig.~\ref{Fig:CoupResHuyCell}(b). In the even mode case, the coupling capacitance has zero contribution due to open circuit termination at the symmetry plane, and the resonator inductance is increased by a factor of $M$. This even mode corresponds to the electric-resonance of the unit cell. On the other hand, for the odd mode case, the coupling capacitor is present accounting for the displacement current between the two metallic patches, and the resonator inductance is reduced by a factor of $M$. This corresponds to the magnetic resonance mode of the unit cell. Each of these circuits are purely reflective type with unity reflectance amplitude (i.e. $|\Gamma_{e,o}(\omega)|=1$), but with frequency-dependent reflection phase, $\angle \Gamma_{e,o}(\omega)$. These two respective reflection coefficients of the even and odd mode circuits are given by

\begin{subequations}
\begin{align}
\Gamma_e &= \left(\frac{Z_{L, \text{even}} - Z/2}{Z_{L, \text{even}} + Z/2}\right)\\
\Gamma_o &= \left(\frac{Z_{L, \text{odd}} - Z/2}{Z_{L, \text{odd}} + Z/2}\right),
\end{align}
\end{subequations}

\noindent where $Z= 377~\Omega$ is the free-space impedance, and the corresponding even and mode impedances are further given by

\begin{subequations}
\begin{align}
Z_{L, \text{even}} &=  \parallelTwo{Z_\text{oc}}{\left[\frac{1}{2j\omega C_0} + j\omega \left(\frac{L_0}{2} + M\right)\right]}\\
Z_{L, \text{odd}} &=  \parallelThree{Z_\text{sc}}{\left[\frac{1}{2j\omega C_0} + j\omega \left(\frac{L_0}{2} - M\right)\right]}{\frac{1}{2j\omega C_m}}
\end{align}
\end{subequations}

\noindent with $Z_\text{oc} = -jZ_0\cot\left(\frac{\beta\ell}{2}\right)$ and $Z_\text{sc} = jZ_0\tan\left(\frac{\beta\ell}{2}\right)$ as the open and short-circuited transmission line impedances, respectively. Finally, the overall reflectance of the unit cell can be constructed using the superposition of even and odd mode reflectances as,

\begin{subequations}
\begin{align}
R(\omega) &=  \left(\frac{\Gamma_e(\omega) + \Gamma_o(\omega)}{2}\right),\\
T(\omega) &=  \left(\frac{\Gamma_e(\omega) - \Gamma_o(\omega)}{2}\right).
\end{align}
\end{subequations}

\noindent From the phenomenological point of view, at any frequency, where the reflection phases of the even and odd modes differ by $180^\circ$, a destructive interference between these modes occur, and a perfect transmission is achieved. These frequency points subsequently correspond to perfect formation of a Huygens' configuration and typically exhibits reflection nulls, similar to the ones observed in Fig.~\ref{Fig:DieelctricHuygensCell}, for instance.

To confirm the Huygens' response of the proposed unit cell and the validity of the equivalent circuit model, Fig.~\ref{Fig:CoupResHuyCell}(c) shows a 
typical full-wave response of the coupled resonator computed using FEM-HFSS for two cases: unbalanced and balanced cases. In the unbalanced case,  the cell exhibits a strong reflectance in a large bandwidth, with two strong transmission nulls. \jn{It indicates a strong disparity between the electric (even) and magnetic (odd) response of the cell. Besides, the transmission phase has two breaking points aligning with those two transmission nulls. On the other hand, for the balanced case, with slight geometrical tuning of the unit cell, a wide-band matching is achieved. It presents two distinct reflectance dips identical to the ones observed in the dielectric cell of Fig.~\ref{Fig:DieelctricHuygensCell}, in addition to a complete $2\pi$ phase coverage.} Remarkably, the proposed equivalent circuit model is able to faithfully reproduce the full response across a large bandwidth for both unbalanced and balanced cases, where the circuit values were extracted using the full-wave results. In both cases, the dielectric impedance and the electrical length \jn{were} fixed to its TEM values. \jn{Large self-inductances $L_0$ are observed in both cases, while very small self-capacitances $C_0$, in the order of fF, indicate a very weak intercellular coupling.} The cell response was further found to be very sensitive to the capacitive and inductive coupling. While no one-to-one mapping was observed between the geometrical parameters and the circuit elements, the self-resonant terms were found to be primarily responsible for the frequency locations of the reflection nulls, and the inductive coupling dominantly affected the matching.

To further understand the Huygens' response of the proposed cell, eigenmode analysis was performed on a practical lossy cell, as shown in Fig.~\ref{Fig:EigenHuygensCell}. Fig.~\ref{Fig:EigenHuygensCell}(a) shows the typical configuration of the unit cell and is typical driven simulation results \jn{under a balanced condition.} The transmission and reflection characteristics show the signature reflection dips corresponding to two Huygens' frequencies. Fig.~\ref{Fig:EigenHuygensCell}(b) further shows the eigenmode field plots for various resonant modes founds within the frequency range of interest. Similar to the dielectric unit cell, again two pairs of electric and magnetic dipole type responses were found near the two reflection nulls. Compared to the dielectric resonators, the dipolar field distributions are less distinct, and large variations in the Q-factors were observed. However, the existence of co-located electric and magnetic dipole pairs further explains the Huygens' configuration in the proposed unit cell structure.

\section{Huygens' Transmit-arrays}

\begin{figure}[htbp]
	\centering
\includegraphics[width=1\columnwidth]{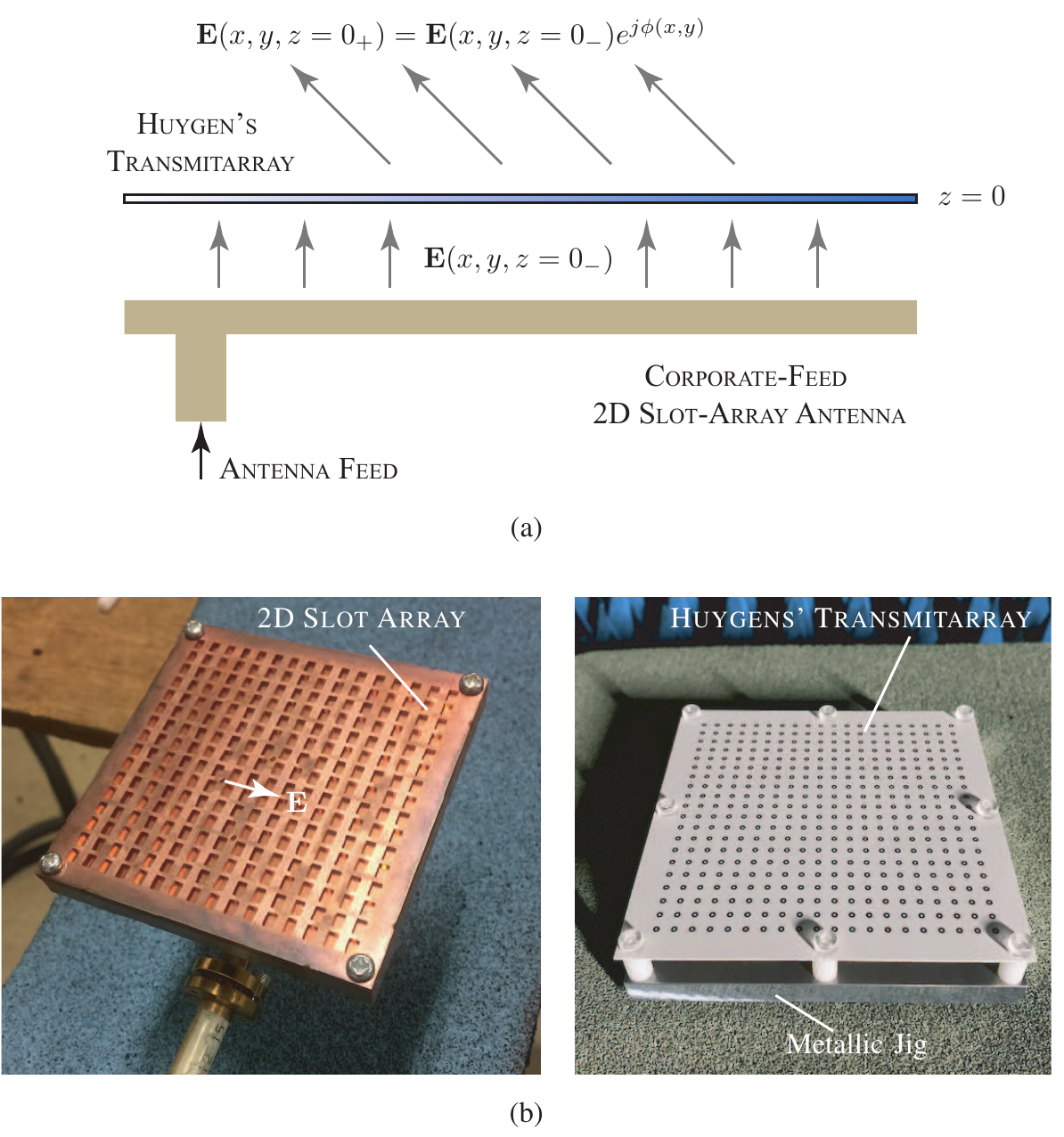}
	\caption{Principle of antenna beam-forming using proposed Huygens' transmit-arrays on top of the 2D Slot array structure. a) 2D slot array antenna only. b) The proposed Huygens' transmit-array (uniform structure as an example) placed on top of the slot array in its near-field.}
	\label{Fig:BFprinciple}
\end{figure}

As demonstrated, the proposed Huygens' unit cell based on coupled resonators provides a wide-band matching. For unit ellipticities, the cell is perfectly identical along $x$ and $y$ axes, which makes it suitable for handling circularly polarized waves. However, to show its beam-forming capabilities, an example of wave polarizations will be taken using a linearly-polarized 2D slot array antenna. The principle of antenna beam-forming using the proposed Huygens' cell is illustrated in Fig.~\ref{Fig:BFprinciple}(a). \jn{It consists of a 2D slot array antenna, with a Huygens' transmit-array placed on top, in its near field.}

The slot array antenna used in this work is a $16\times 16$ array of radiating slots forming a 2D linearly polarized structure, fabricated using the diffusion bonding process, as shown in Fig.~\ref{Fig:BFprinciple}(b). It is a corporate feed antenna with a standard waveguide input at the bottom. \jn{The transmit-array on top of the antenna aperture should ideally provide a uniform transmission amplitude along with space-varying phase distribution. Thus, transforming an incoming ideal plane wave $\mathbf{E}(x,y, z=0_-)$ to a desired outgoing wave $\mathbf{E}(x,y, z=0_+)$ at the design frequency.} It specifically imparts a phase of $\phi(x,y)$ which is engineered to achieve a desired far-field radiation characteristic. A photograph of a typical Huygens' transmit-array is shown in Fig.~\ref{Fig:BFprinciple}(b), which is placed in the near-field of the slot array using a metallic jig to support the structure with nylon cylindrical posts.

\begin{figure}[htbp]
\begin{center}
\includegraphics[width=1\columnwidth]{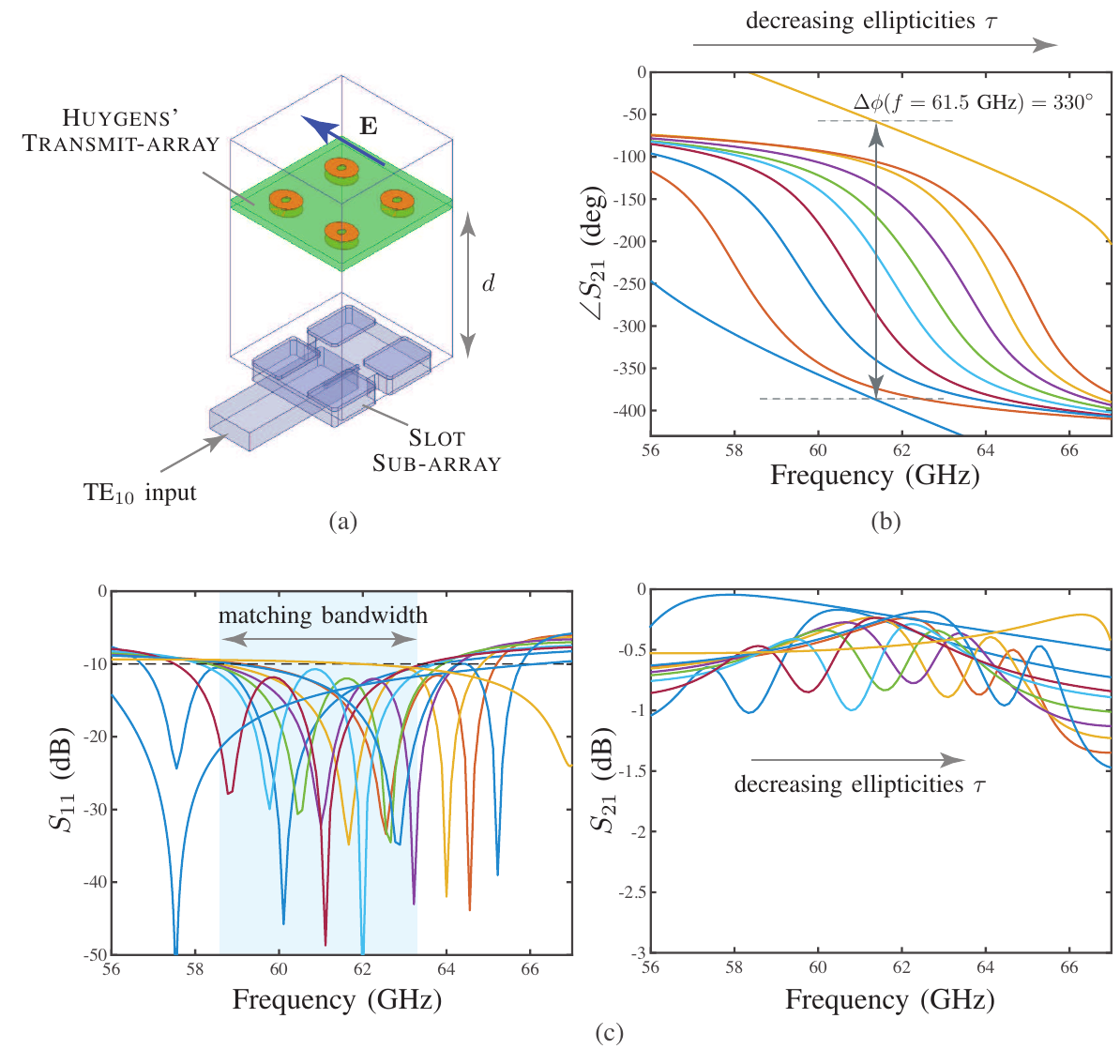}
 \caption{FEM-HFSS simulation setup of the Huygens' cell on top of the slot sub-array. a) Setup. b) Transmission phase tuning by varying cell geometry. c) Corresponding transmission and reflection response. Simulation parameters are: $\epsilon_r = 2.00$, $\tan\delta = 0.0021$, substrate thickness $h=0.508$~mm, unit cell size $\Lambda = 4.2$~mm, $\tau_\text{out} \in (0.1,~4.0),~a_\text{out} \in (500,~1100)~\mu$m and $\tau_\text{out} \in (0.4,~2.3),~a_\text{out} \in (10,~275)~\mu$m.}\label{Fig:FEMSingleHuygensCell}
\end{center}
\end{figure}

While the Huygens' transmit-array is expected to have zero back-reflection, the typical reflection characteristics of the proposed unit cell exhibits finite amount of reflection, as seen in Fig.~\ref{Fig:EigenHuygensCell}. To take into account any possible coupling between the slot elements and the transmit-array cells, a compound structure is used in the design as shown in Fig.~\ref{Fig:FEMSingleHuygensCell}(a). It shows a $2\times2$ slot sub-array fed by a standard TE$_{10}$ waveguide, which is the building block of the larger array. The Huygens' cell period is matched to the slot subarray period for convenient unit cell design with $\Lambda=4.2$~mm, so that there is one \jn{Huygens element at the top of each slot}. The separation distance $d$ between the slots and the transmit-array is chosen to minimize the loading effect of the transmit-array on the slot, as the slot array antenna is assumed to be fixed once fabricated. All vertical air faces are assigned Floquet boundaries to emulate a uniform field distribution, with radiation boundaries on top. \blue{By varying the unit cell ellipticities, and adjusting the matching using $a_\text{in}$ and $a_\text{out}$, the transmission phase is varied across frequency, while maintaining broadband matching.} The transmission phase variation is shown in Fig.~\ref{Fig:FEMSingleHuygensCell}(b) and the corresponding reflection and transmission characteristics are shown in Fig.~\ref{Fig:FEMSingleHuygensCell}(c). A wide-band matching is observed in each case, typically with $|S_{11}| < -10$~dB. The transmission losses are kept below $-1$~dB within the desired bandwidth. \jn{The maximum phase variation at a fixed frequency is found to be about $330^\circ$ at 61.5~GHz. For these design parameters, the setup maintains a good reflection performance in a large absolute bandwidth of about 4.25~GHz.}

\begin{figure}[htbp]
\begin{center}
\includegraphics[width=1\columnwidth]{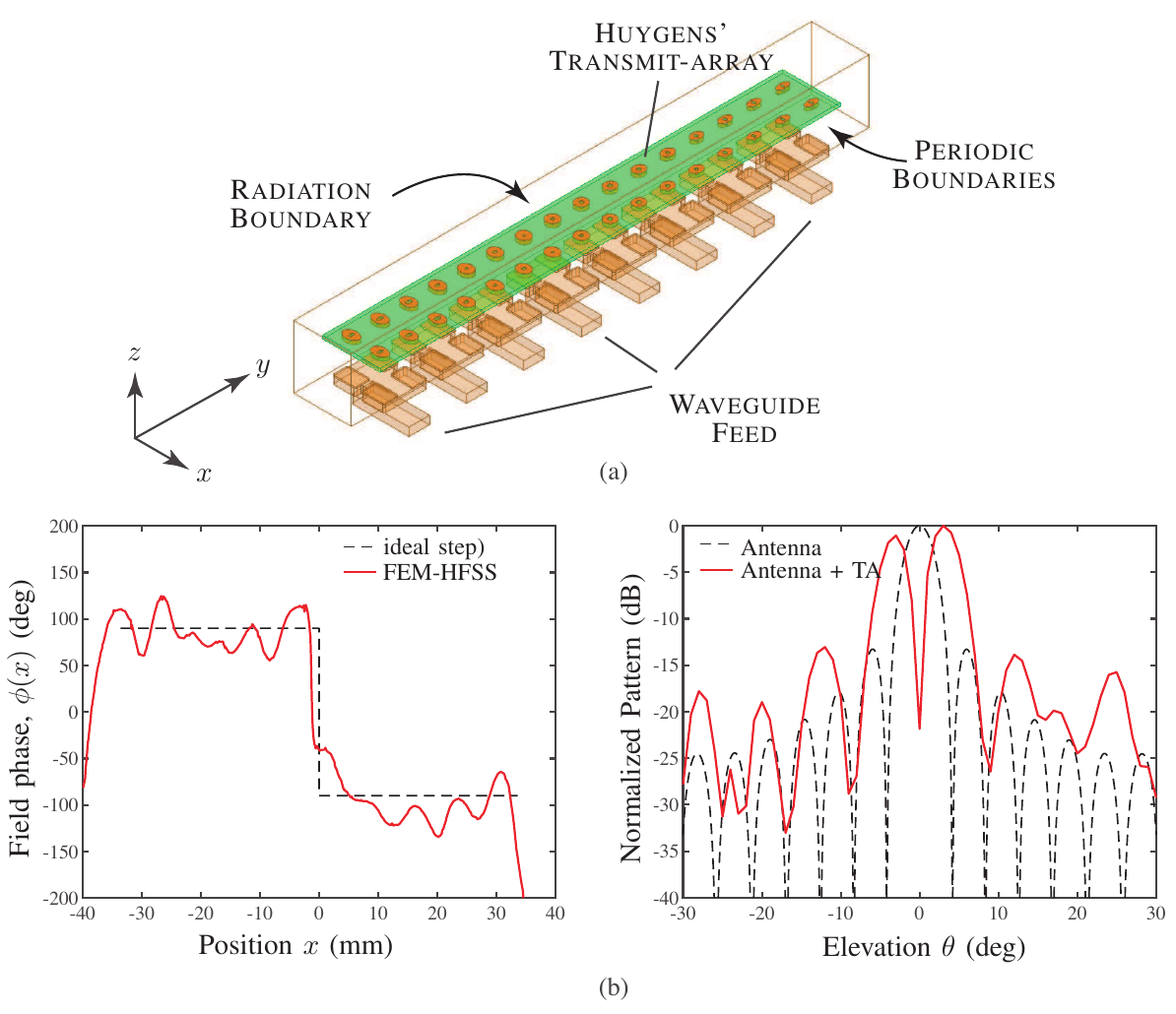}
 \caption{FEM-HFSS simulation setup of a typical Huygens' transmit-array on top of the slot array antenna \jn{to demonstrate its phase} engineering capability using a difference pattern generation as an example. a) Simulation setup. b) Near-field phase at the design frequency, \jn{and (c) }the corresponding far-field radiation pattern of the compound structure compared to the antenna only. Simulation parameters are provided in Table.~I.}\label{Fig:FEMTAHuygensCell}
\end{center}
\end{figure}

\begin{figure*}[htbp]
\begin{center}
\includegraphics[width=2\columnwidth]{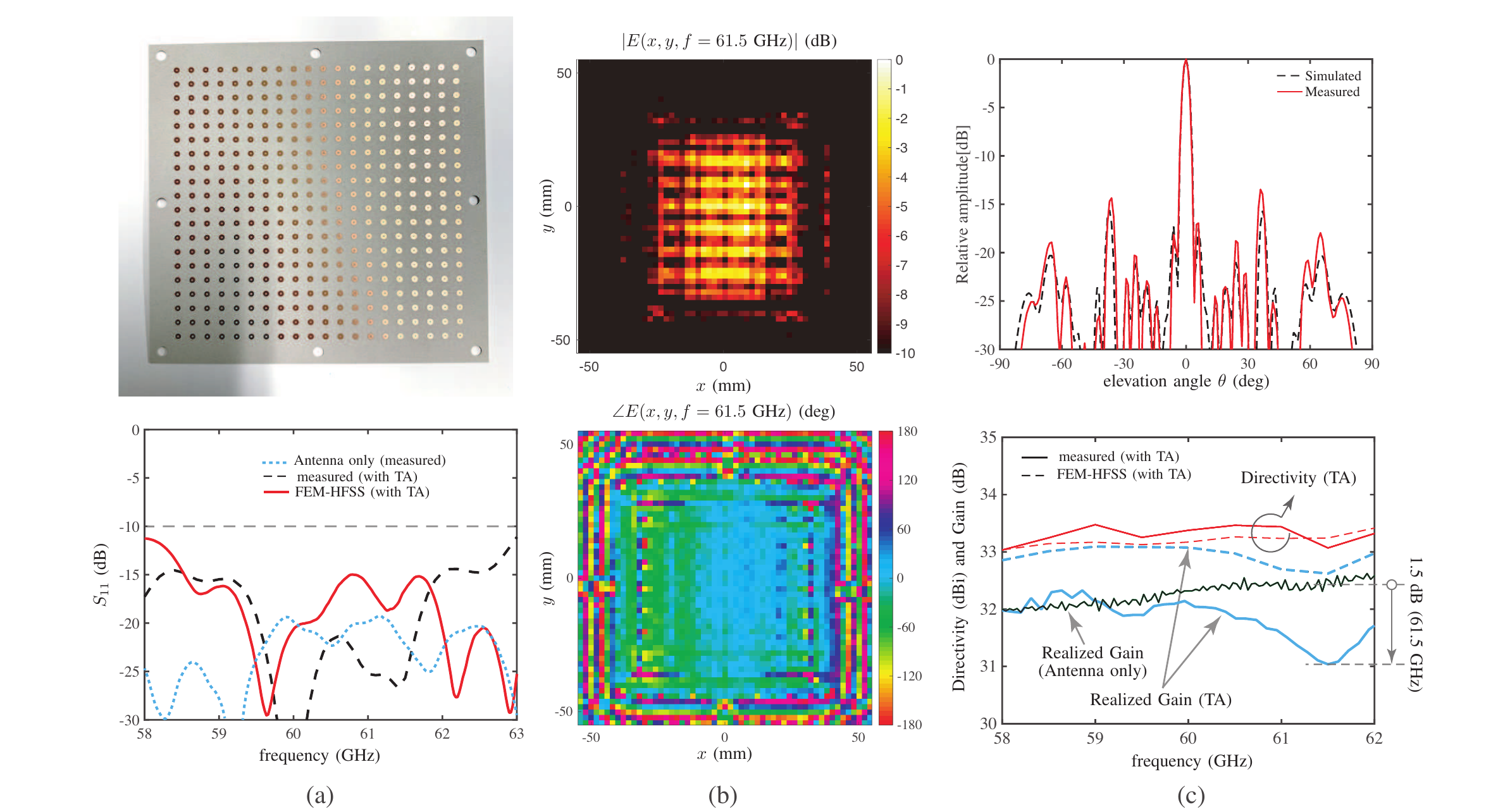}
 \caption{Measured reflection and field characteristics of a uniform transmit-array. a) Picture of the transmit-array and the measured broadband reflection characteristics of the antenna with and without the Huygens' transmit-array. b) 2D near-field amplitude and phase response at the design frequency. c) Far-field radiation pattern, gain and directivity (solid curve - measurements, dashed curves - FEM-HFSS) of the antenna with the transmit-array (TA).}\label{Fig:Uniform}
\end{center}
\end{figure*}

As an illustrative beam-forming example, consider the case of a difference pattern generation, where two beams are desired with a null at broadside \cite{Sherman_Monopulse}. Such a transmit-array is realized by dividing the array region into two halves with $180^\circ$ transmission phase difference at the design frequency. The simulation setup of general case of 1D transmit-array is illustrated in Fig.~\ref{Fig:FEMTAHuygensCell}(a). It consists of a 1D array of 8 slot sub-arrays (total 16 slot elements) with Huygens' transmit-array on top, separated by a distance $d$ from the antenna aperture. Floquet boundaries are applied on all four vertical sides of the air box with radiation boundaries on top. For difference pattern generation, the geometrical parameters of the Huygens' transmit array is varied in a desired fashion, and the corresponding spatial phase distribution is shown in Fig.~\ref{Fig:FEMTAHuygensCell}(b), measured on top of the surface. \jn{Around 180$^\circ$ phase difference is achieved at 61.5~GHz, with slight ripples on each side.} The corresponding far-field radiation pattern is also shown in Fig.~\ref{Fig:FEMTAHuygensCell}(b). Compared to the broadside radiation of the original antenna, the transmit-array successfully generates a strong null at broadside and beams on either side. With the choice of the optimal value of $d = 0.5$~cm, all the waveguide input ports are very well matched, well below $-10$~dB. It should be noted that, while the individual unit cells are designed within a perfectly periodic environment, the transit-array is in general, non-uniform in nature, which is expected to lead to some discrepancies when using the unit cell design. Using this numerical setup, various transmit-array prototypes were fabricated and tested, as shown next.

\begin{figure*}[htbp]
\begin{center}
\includegraphics[width=2\columnwidth]{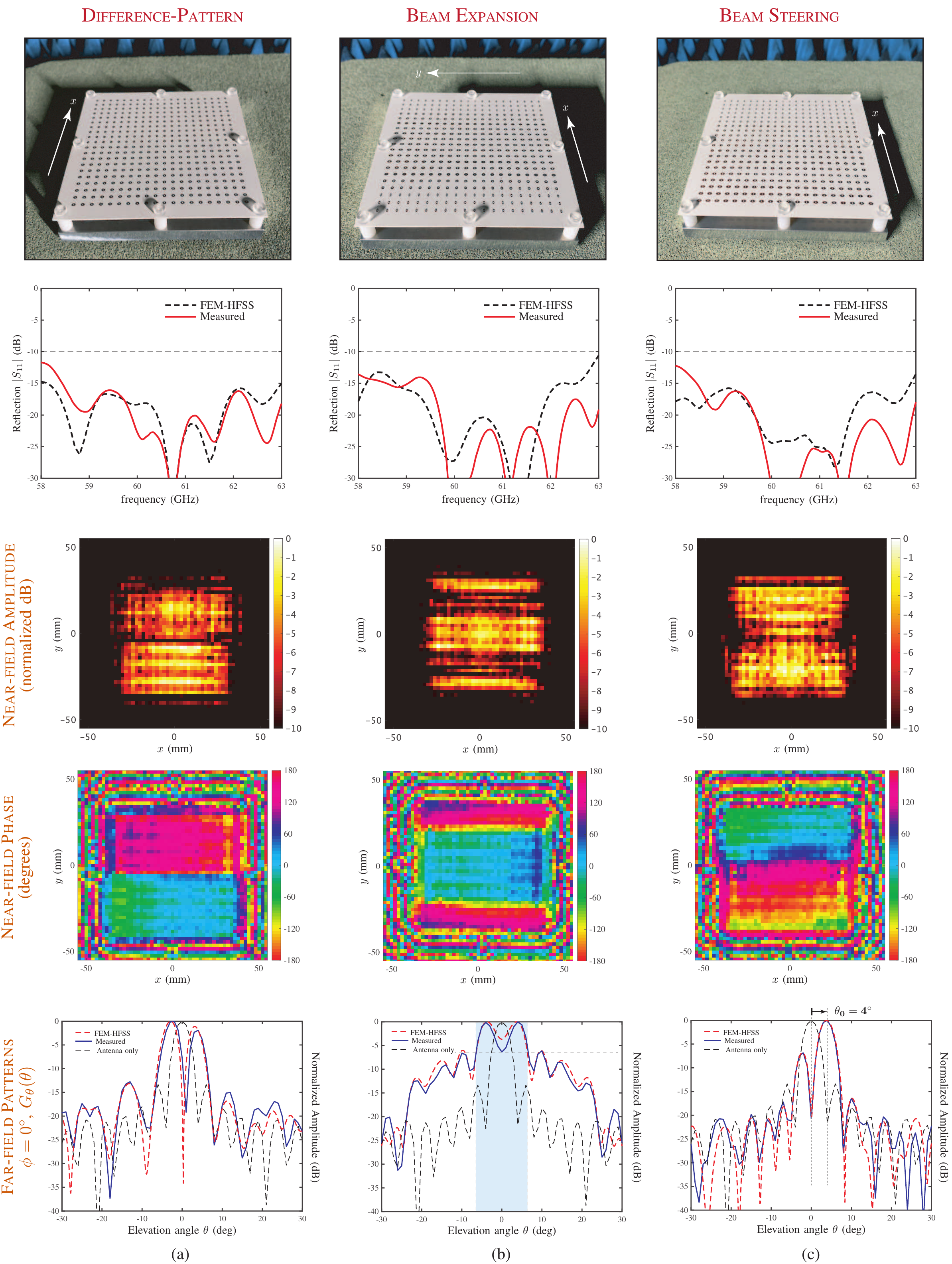}
 \caption{Measurement results of three fabricated transmit-arrays with different phase profiles at 61.5~GHz, showing the photograph of the prototypes, reflection characteristics, near-field aperture fields and the far-fields.}\label{Fig:Prototypes}
\end{center}
\end{figure*}

\section{Experimental Demonstration \& Characterization}

To experimentally characterize the proposed Huygens' transmit array, four different prototypes were made and tested: 1) Uniform surface, 2) Difference pattern generation, 3) Beam expansion and 4) Beam steering. The $16\times16$ linearly polarized slot array antenna is kept the same in all cases, and the transmit-array is mounted on a metallic jig as shown in Fig.~\ref{Fig:BFprinciple}(b). The substrate has a permittivity of $\epsilon_r =2$ and $h=0.508$~mm, which is $\approx \lambda/10$ thick at 60~GHz. 

Let us consider the first prototype of uniform surface, which is useful in estimating the typical loss characteristics of the surface. The picture of the fabrication prototype is shown in Fig.~\ref{Fig:Uniform}(a). The figure also shows the measured reflection, which is found to be well below  $-10$~dB in a large bandwidth between 58-63 GHz. The near-fields on top of the transmit-array were next measured and the 2D field magnitude and phase is shown in Fig.~\ref{Fig:Uniform}(b). A perfectly uniform amplitude and phase is desired throughout the surface. However, the near-field shows that the field amplitude decreases near the edge of the surface, which can be attributed to the edge effects of the finite sized array. Moreover, there is a gradual phase shift observed along the $x-$direction which is an indicative of the slight vertical misalignment of the surface with respect to the antenna aperture. The corresponding far-field patterns are shown in Fig.~\ref{Fig:Uniform}(c), with peak radiation still along broadside indicating negligible effect of the slight phase gradient in the near-fields, on the far-fields. The corresponding directivity and measured gain is next shown in Fig.~\ref{Fig:Uniform}(c) with a peak directivity of about 33~dBi throughout the frequency band, where the directivity is estimated using numerical far-field transformation of the measured near-fields. A good agreement is seen between FEM-HFSS and measurement. On the other hand, the measured realized gain is found to be lower than the simulated values by about 1.56~dB (61.5~GHz), indicating under-estimated values of copper conductivity and loss tangent of the dielectric substrate in the design simulations. Comparison with the measured realized gain of the antenna only, reveals a maximum gain drop of 1.5~dB at 61.5~GHz, which is a representative of the actual loss due to the Huygens' transmit-array (considering negligible reflection losses, and assuming same directivities).

\begin{table*}[t]  \caption{Geometrical design parameters for various prototypes shown in Fig.~\ref{Fig:Uniform} and Fig.~\ref{Fig:Prototypes}.} \label{tab:1}
  \centering
  \begin{tabular}{cccccccccccccccccc}\toprule
    \multicolumn{17}{c}{\textbf{Beam Expansion}} & \multirow{2}{*}{\textbf{Uniform}} \\ \cmidrule{2-17}
     & \#1 & \#2 & \#3 & \#4 & \#5 & \#6 & \#7 & \#8 & \#9 & \#10 & \#11 & \#12 & \#13& \#14 & \#15 & \#16 &  \\ \midrule
    $a_\text{in}$ & 400 & 440 & 275 & 350 & 130 & 275 & 200 & 190 & 190 & 200 & 275 & 130 & 350 & 275 & 440 & 400 & 275 \\
    $\tau_\text{in}$ & 0.5 & 0.54 & 0.61 & 0.72 & 0.86 & 1.04 & 1.25 & 1.5 & 1.5 & 1.25 & 1.04 & 0.86 & 0.72 & 0.61 & 0.54 & 0.5 & 1.03 \\
    $a_\text{out}$ & 775 & 775 & 800 & 800 & 850 & 850 & 900 & 900 & 900 & 900 & 850 & 850 & 800 & 800 & 775 & 775 & 850\\
    $\tau_\text{out}$ & 1.5 & 1.46 & 1.39 & 1.28 & 1.14 & 0.96 & 0.75 & 0.5 & 0.5 &0.75 & 0.96 & 1.14 & 1.28 & 1.39 & 1.46 & 1.5 & 0.97 \\ \toprule
    \multicolumn{17}{c}{\textbf{Beam Steering}} & \multirow{2}{*}{\textbf{Difference-Pattern}} \\ \cmidrule{2-17}
     & \#1 & \#2 & \#3 & \#4 & \#5 & \#6 & \#7 & \#8 & \#9 & \#10 & \#11 & \#12 & \#13& \#14 & \#15 & \#16 &   \\ \midrule
    $a_\text{in}$ & 400 & 440 & 450 & 350 & 250 & 285 & 200 & 260 & 275 & 290 & 240 & 200 & 250 & 225 & 220 & 190 & 285/200\\
    $\tau_\text{in}$ & 0.5 & 0.57 & 0.63 & 0.70 & 0.77 & 0.83 & 0.90 & 0.97 & 1.03 & 1.10 & 1.17 & 1.23 & 1.3 & 1.37 & 1.43 & 1.5 & 0.83/1.23\\
    $a_\text{out}$ & 5 & 6 & 7 & 8 & 5 & 6 & 7 & 8 & 5 & 6 & 7 & 8 & 5 & 6 & 7 & 8 & 825/900\\
    $\tau_\text{out}$ & 775 & 775 & 775 & 800 & 825 & 825 & 850 & 850 & 850 & 850 & 875 & 900 & 875 & 900 & 900 & 925 & 1.17/0.77 \\ \bottomrule
    \multicolumn{18}{l}{\footnotesize All dimensions in $\mu$m, $\epsilon_r = 2.00$, $\tan\delta = 0.0021$, substrate thickness $h=0.508$~mm and unit cell size $\Lambda = 4.2$~mm.}
  \end{tabular}
\end{table*}

Next, three different prototypes were fabricated corresponding to difference-pattern generation, beam expansion and beam-steering, as shown in Fig.~\ref{Fig:Prototypes}. The difference-pattern generation creates a strong null along broadside, where there is a phase difference of $180^\circ$ between transmission phase of left and half regions of the transmit-array, leading to destructive interference. Such a beam is useful in creating two regions of illumination and avoiding region in between, in the far-field, for instance in the GATE application. Fig.~\ref{Fig:Prototypes}(a), shows the picture of the prototype (transmit-array on top of the slot array antenna), along with its measured reflection characteristics. Broadband matching is achieved as expected and showing good comparison with FEM-HFSS design, confirming small fabrication tolerances. The near-field amplitude and phase scan at the chosen frequency of 61.5~GHz, clearly show the separation of the two regions on the surface, with $\approx 180^\circ$ phase difference. There is a slight phase gradient seen along the $x-$direction, which is attributed to possible vertical misalignment between the slot array plane and the transmit-array plane, as the surface is uniform along this direction. The corresponding far-field patterns produce the desired nulls with about $1.5$~dB difference in the two peak values of the beam.

Second prototype for beam expansion is shown in Fig.~\ref{Fig:Prototypes}(b). This surface exhibits a concave quadratic phase profile along the $x-$direction. Similar to a cylindrical concave lens, this transmit-array is useful in broadening the far-field pattern providing larger coverage area in the GATE application, for instance. Fig.~\ref{Fig:Prototypes}(a) again shown good wide-band matching performance with $|S_{11}| < -10$~dB. The near-field scan clearly shows the concave phase profile with minimum phase in the central region and quadratically increasing on either sides until the edge of the structure. While an ideally flat magnitude response is desired, different cells with different transmission phases also exhibit different transmission magnitudes, as shown in Fig.~\ref{Fig:FEMSingleHuygensCell}. The far-field patterns shows a significant beam expansion compared to a standalone pattern, with good matching with the numerical design. A 6~dB beam width of about 13$^\circ$ is observed compared to only about $5^\circ$ for the antenna.

The last prototype is a beam-steering structure, as shown in Fig.~\ref{Fig:Prototypes}(c). It consists of linearly varying phase across $x-$axis, creating a constant non-zero phase gradient, required for steering the main-beam of the antenna. The desired phase gradient is $\phi_x = 2\pi/(16\times 4.2)$~rad/mm. This corresponds to a beam tilt of $\theta_0 = \sin^{-1}\{\phi_x/k_0\}$, where $k_0$ is the wave number at the design frequency. This leads to a theoretical beam tilt of about $4.16^\circ$ in the far-field at 61.5~GHz. The near-fields of Fig.~\ref{Fig:Prototypes}(c) shows the linear phase gradient along the $x-$direction, where again a non-negligible phase gradient is observed along the $y-$direction. Nevertheless, the far-field patterns shows a clear tilt of $4^\circ$ in very good agreement with the theoretical value.

All these examples, show a glimpse of the diversity of phase profiles that may be achieved using the proposed Huygens' transmit-array with wide-band matching bandwidths. Based on standard and simple PCB processes with no via connections, all the prototypes exhibited negligible fabrication tolerances. All geometrical dimensions of the four prototypes are finally tabulated in Tab.~I for reference.

\section{Discussions}

\subsection{Unit Cell Period vs Bandwidth Trade-offs}

While the proposed coupled resonator configuration provides a large phase range, the spatially varying phase profiles are limited primarily due to relatively large unit cell sizes, similar to that of dielectric-based Huygens' surfaces. The cell period, $\Lambda$ in all the previous designs has been chosen as 4.2~mm to match with that of the slot period of an already available antenna prototype ($\approx 0.86\lambda_0$ at 61.5~GHz). This led to simpler designs of the transmit-array, as the coupling between slot sub-array and the surface can be efficiently modeled.

\begin{figure}[htbp]
\begin{center}
\includegraphics[width=1\columnwidth]{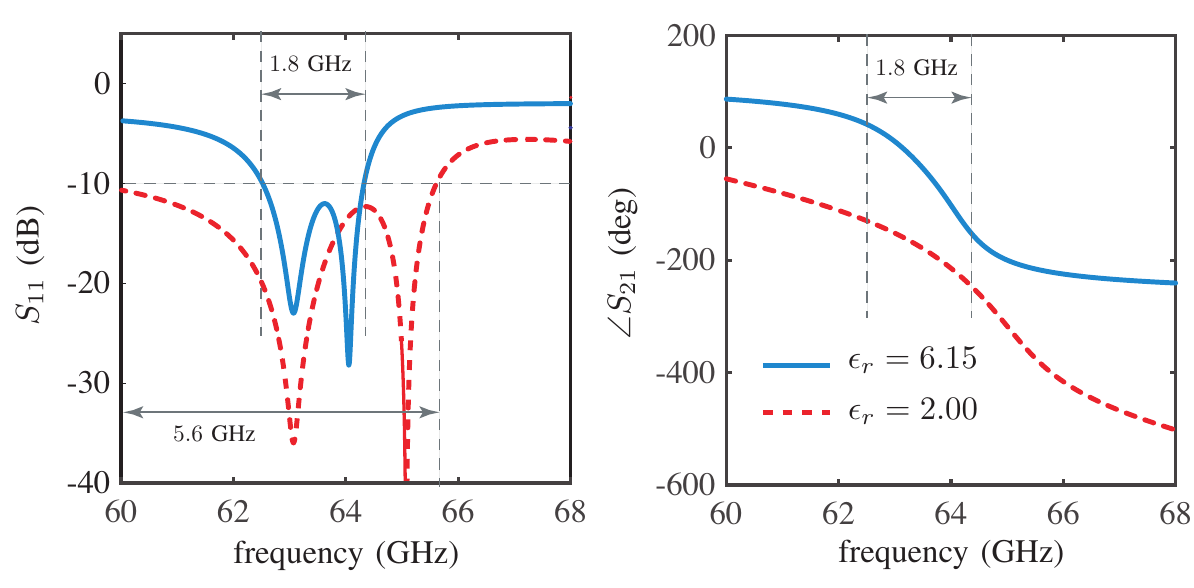}
\caption{Trade-off between unit cell size and matching bandwidth. Simulation parameters are: $\epsilon_r = 6.15$, $\tan\delta = 0.0025$, $\tau_\text{out} = 0.6,~a_\text{out} = 525~\mu$m, $\tau_\text{in} = 1.0,~a_\text{in} = 100~\mu$m, substrate thickness $h=0.35$~mm and $\Lambda = 2.8$~mm. The reference design parameters are the same as in Fig.~\ref{Fig:EigenHuygensCell}.}\label{Fig:SmallCells}
\end{center}
\end{figure}

However, the unit cell period may be reduced to enable better spatial phase discretization. One simple way is by using a higher permittivity dielectric between the two coupled resonators. Fig.~\ref{Fig:SmallCells} shows one example, where a substrate of $\epsilon_r = 6.15$ (Rogers 3006) is used to design the Huygens' cell. Using the same elliptically shaped metallic resonator, the unit cell has been reduced to $\Lambda = 2.8$~mm, which is about $0.57\lambda_0$ at 61.5~GHz. However, this led to reduced matching bandwidth compared to the design based on $\epsilon_r = 2.00$, which is only about 32\% of the original design (1.8~GHz vs 5.6~GHz). Nevertheless, the transmission phase shows a faster frequency variation, which indicates a larger sensitivity to resonator dimensions. This example thus highlights the fundamental inherent trade-off between higher permittivity based smaller unit cell periods and the lower matching bandwidths in the proposed Huygens' cell. 

\subsection{Circular Polarization Handling}

\begin{figure}[htbp]
\begin{center}   
\includegraphics[width=1\columnwidth]{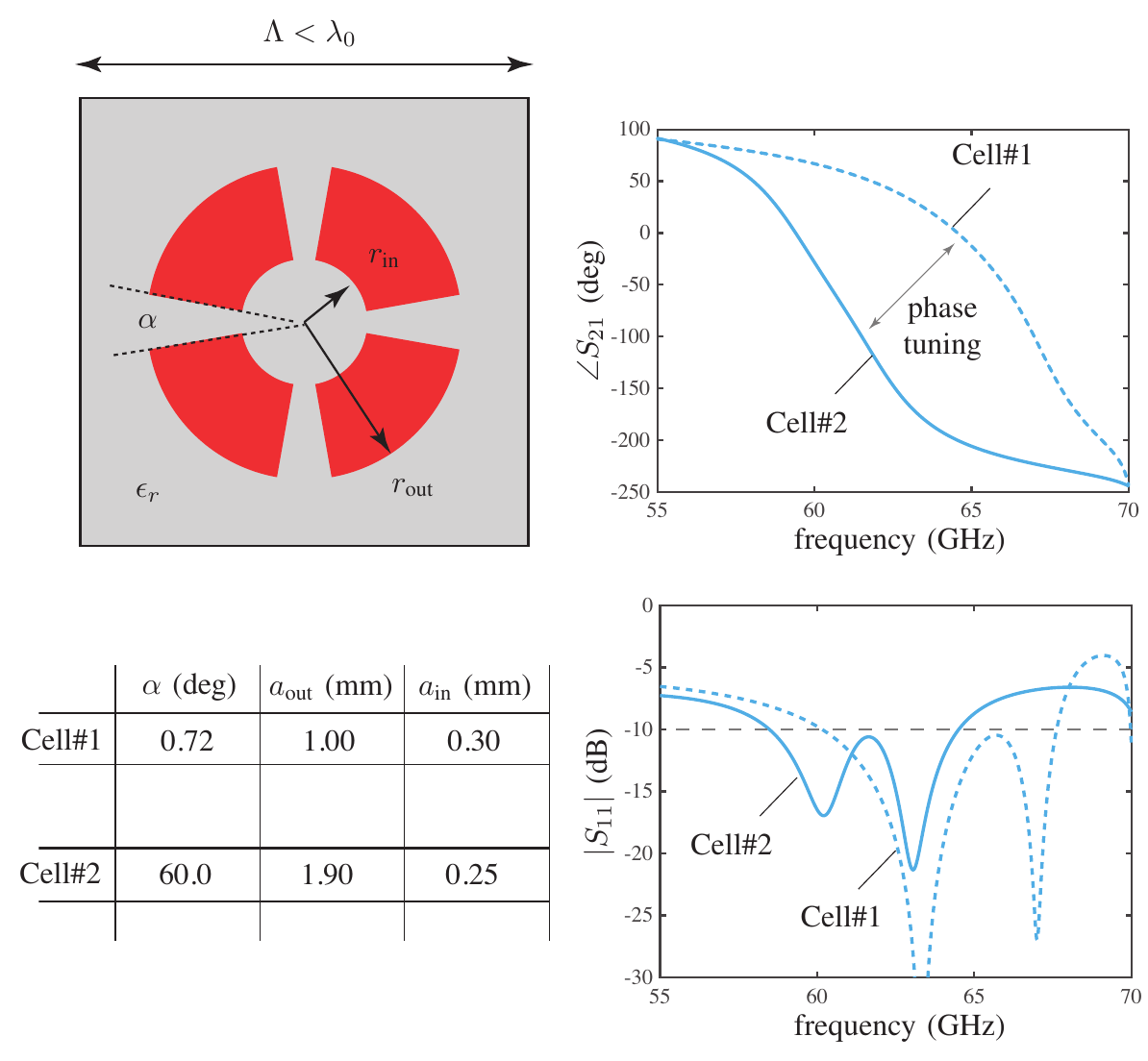}
\caption{Extension of the proposed Huygens' cell to handle circularly polarized waves enforcing rotational symmetry in the unit cell design. $\epsilon_r = 2.00$, $\tan\delta = 0.0025$, substrate thickness $h=0.538$~mm and $\Lambda = 4.2$~mm.}\label{Fig:CPCells}
\end{center}
\end{figure}

The proposed cell has been demonstrated for handling linear polarization. However, there are applications such as GATE systems, which are based on circularly polarized slot array antennas. The proposed Huygens' transmit-array configuration may also be extended to operate on circular polarization as well, with some modifications in the geometry. Fig.~\ref{Fig:CPCells} shows one illustrative example where the metallic resonator patches have been modified to operate on circular polarization. Consequently, a rotational symmetry has been imposed on the metallic resonator. Since some design parameters are lost, because $\tau_\text{out} = \tau_\text{in} = 1$, a new geometrical parameter is introduced as an annular slit of angle $\alpha$ on the metallic ring resonator. Fig.~\ref{Fig:CPCells} shows the transmission phase for two sets of parameters, where the Huygens' response with two distinct reflection nulls is clearly visible, and its phase response is varied across frequency. Due to the imposed symmetry, both orthogonal polarizations naturally have the same response, as desired for circular polarization operation. While this serves as a simple example to illustrate possible extensions to circular polarized waves, more sophisticated geometrical modifications may be needed to enable greater control over its scattering response.

\section{Conclusions}

A novel Huygens' transmit-array has been proposed compatible with standard PCB processes and a single dielectric substrate at mm-Waves for beam-forming application. It has been constructed using a coupled-resonator approach where two identical elliptical metallic patches with an elliptical hole separated by a dielectric substrate has been used. By engineering the dimensions of the resonator, the electric and magnetic resonances have been excited to achieve zero back-scattering in a large bandwidth. The operation principle of the proposed Huygens' cell has been explained in details using both an equivalent circuit model, as well as eigenmode analysis, and compared with standard planar all-dielectric Huygens' structures. Finally, the proposed Huygens' cell is placed on top of a high gain slot-array antenna, in its near-field to engineer its aperture field distribution, acting as a phase plate. Several transmit-array prototypes have been demonstrated and experimentally characterized in both their near and far-fields, corresponding to pattern generation, beam expansion and beam steering as application examples, in addition to a uniform surface demonstrating low-loss performance around 60~GHz frequency band. Further discussions related to unit cell size vs frequency bandwidth trade-offs and possible extension to handling circular polarization have been finally provided.

\jn{The proposed single dielectric coupled resonator Huygens' structure has been classified here as a transmit-array due to its large unit cell size. However, it is anticipated that with even larger dielectric constant substrates, the unit cell size can be further reduced. Thus, qualifying the structure as an electromagnetic metasurface with deeply sub-wavelength periodicities. This will greatly enhance its beam forming capabilities beyond transmit-array operation, by enabling larger spatial phase variations at the cost of smaller matching bandwidths.} Besides, the near-field operation of the transmit-arrays makes the overall compound structure compact in size, where the transmit-array may also be regarded as an electromagnetically active radome to the slot array antenna structure. Future works include a rigorous mapping between the equivalent circuit elements and the geometry of the unit cell which may lead to faster unit designs along with deeper insights to unlock real-time phase tuning capabilities. Extensions to handling circular polarizations and \jn{their detailed experimental characterization} is another interesting avenue to pursue. Thus the proposed Huygens' transmit-array may represent an important step forward in implementing high efficiency electromagnetic structures for both passive and active beam-forming in the mm-Wave frequency region for next generation wireless systems.

\bibliographystyle{IEEEtran}
\bibliography{2020_PCB_mmWAVE_METASURFACE_TAP_SAKURAI}

\begin{thebibliography}{10}
\providecommand{\url}[1]{#1}
\def\UrlFont{\rmfamily}
\providecommand{\newblock}{\relax}
\providecommand{\bibinfo}[2]{#2}
\providecommand\BIBentrySTDinterwordspacing{\spaceskip=0pt\relax}
\providecommand\BIBentryALTinterwordstretchfactor{4}
\providecommand\BIBentryALTinterwordspacing{\spaceskip=\fontdimen2\font plus
\BIBentryALTinterwordstretchfactor\fontdimen3\font minus
  \fontdimen4\font\relax}
\providecommand\BIBforeignlanguage[2]{{%
\expandafter\ifx\csname l@#1\endcsname\relax
\typeout{** WARNING: IEEEtran.bst: No hyphenation pattern has been}%
\typeout{** loaded for the language `#1'. Using the pattern for}%
\typeout{** the default language instead.}%
\else
\language=\csname l@#1\endcsname
\fi
#2}}

\bibitem{Holloway_Metasurface}
C.~L. {Holloway}, E.~F. {Kuester}, J.~A. {Gordon}, J.~{O'Hara}, J.~{Booth}, and
  D.~R. {Smith}, ``An overview of the theory and applications of metasurfaces:
  The two-dimensional equivalents of metamaterials,'' \emph{IEEE Antennas
  Propag. Mag.}, vol.~54, no.~2, pp. 10--35, Apr. 2012.

\bibitem{Tretyakov_MS}
S.~A. {Tretyakov}, ``{A personal view on the origins and developments of the
  metamaterial concept},'' \emph{J. Opt.}, vol.~19, no.~1, p. 013002, Jan 2017.

\bibitem{Tretyakov_Huygens}
T.~S. A., ``Metasurfaces for general transformations of electromagnetic
  fields,'' \emph{Philos. Trans. Roy. Soc. London A, Math. Phys. Sci.}, vol.
  373, no. 2049, Feb. 2015.

\bibitem{Yu_PhaseControl}
N.~Yu and F.~Capasso, ``Flat optics with designer metasurfaces,'' \emph{Nat.
  Mater.}, vol.~13, no. 139, 2014.

\bibitem{Glybovski_PhaseControl}
S.~B. Glybovski, S.~A. Tretyakov, P.~A. Belov, Y.~S. Kivshar, and C.~R.
  Simovski, ``Metasurfaces: From microwaves to visible,'' \emph{Phys. Rep.},
  vol. 634, pp. 1--72, 2016.

\bibitem{Gharghi_Cloak}
M.~Gharghi, C.~Gladden, T.~Zentgraf, Y.~Liu, X.~Yin, J.~Valentine, and
  X.~Zhang, ``A carpet cloak for visible light,'' \emph{Nano Lett.}, vol.~11,
  no.~7, pp. 2825--2828, 2011.

\bibitem{Moitra_Reflection}
P.~Moitra, B.~A. Slovick, Z.~Gang~Yu, S.~Krishnamurthy, and J.~Valentine,
  ``Experimental demonstration of a broadband all-dielectric metamaterial
  perfect reflector,'' \emph{Appl. Phys. Lett.}, vol. 104, no.~17, p. 171102,
  2014.

\bibitem{BBParticlesTratyakov}
V.~S. Asadchy, I.~A. Faniayeu, Y.~Ra'di, S.~A. Khakhomov, I.~V. Semchenko, and
  S.~A. Tretyakov, ``Broadband reflectionless metasheets: Frequency-selective
  transmission and perfect absorption,'' \emph{Phys. Rev. X}, vol.~5, p.
  031005, Jul 2015.

\bibitem{Shi_PolarizationControl}
H.~Shi, A.~Zhang, S.~Zheng, J.~Li, and Y.~Jiang, ``Dual-band polarization angle
  independent 90$^\circ$ polarization rotator using twisted
  electric-field-coupled resonators,'' \emph{Appl. Phys. Lett.}, vol. 104,
  no.~3, p. 034102, 2014.

\bibitem{Grbic_Metasurfaces}
C.~Pfeiffer and A.~Grbic, ``Metamaterial {H}uygens' surfaces: Tailoring wave
  fronts with reflectionless sheets,'' \emph{Phys. Rev. Lett.}, vol. 110, p.
  197401, May 2013.

\bibitem{MetaHolo}
G.~Zheng, H.~Muhlenbernd, M.~Kenney, G.~Li, T.~Zentgraf, and S.~Zhang,
  ``Metasurface holograms reaching 80\% efficiency,'' \emph{Nat. Nanotech.},
  no.~43, pp. 308--312, Feb. 2015.

\bibitem{Tech_5G_Challenge}
W.~Chin, Z.~Fan, and T.~R. Haines, ``Emerging technologies and research
  challenges for 5g wireless networks,'' \emph{Toshiba Research Europe Limited,
  Bristol, UK}.

\bibitem{5G_mmWave}
T.~S. {Rappaport}, Y.~{Xing}, G.~R. {MacCartney}, A.~F. {Molisch},
  E.~{Mellios}, and J.~{Zhang}, ``Overview of millimeter wave communications
  for fifth-generation (5g) wireless networks—with a focus on propagation
  models,'' \emph{{IEEE Trans. Antennas Propagat.}}, vol.~65, no.~12, pp.
  6213--6230, 2017.

\bibitem{60WaveTech}
``The 60 ghz band cuts 4g backhaul costs,'' \emph{Electronic Design Article},
  2013.

\bibitem{mmWaveTech}
``Millimeter waves: Emerging markets brochure,'' \emph{A Thintri Market Study},
  2012.

\bibitem{GATE}
M.~{Zhang}, K.~{Toyosaki}, J.~{Hirokawa}, M.~{Ando}, T.~{Taniguchi}, and
  M.~{Noda}, ``A 60-{GH}z band compact-range gigabit wireless access system
  using large array antennas,'' \emph{IEEE Trans. Antennas Propag}, vol.~63,
  no.~8, pp. 3432--3440, Aug 2015.

\bibitem{Tomura_Dif_Antenna1}
T.~Tomura, J.~Hirokawa, T.~Hirano, and M.~Ando, ``A 45$^{\circ}$linearly
  polarized hollow-waveguide 16$\,\times$16-slot array antenna covering 71-86
  {GH}z band,'' \emph{IEEE Trans. Antennas Propagat.}, vol.~62, no.~10, pp.
  5061--5067, Oct 2014.

\bibitem{Tomura_Corporate}
T.~Tomura, Y.~Miura, M.~Zhang, J.~Hirokawa, and M.~Ando, ``A 45$^\circ$linearly
  polarized hollow-waveguide corporate-feed slot array antenna in the 60-{GH}z
  band,'' \emph{IEEE Trans. Antennas Propagat.}, vol.~60, no.~8, pp.
  3640--3646, Aug 2012.

\bibitem{DiffusionBond}
M.~{Sano}, J.~{Hirokawa}, and M.~{Ando}, ``A hollow rectangular coaxial line
  for slot array applications fabricated by diffusion bonding of laminated thin
  metal plates,'' \emph{IEEE Trans. Antennas Propag}, vol.~61, no.~4, pp.
  1810--1815, April 2013.

\bibitem{Elef_LWA_MS}
A.~Mehdipour, J.~W. Wong, and G.~V. Eleftheriades, ``Beam-squinting reduction
  of leaky-wave antennas using {H}uygens metasurfaces,'' \emph{{IEEE Trans.
  Antennas Propagat.}}, vol.~63, no.~3, pp. 978--992, Mar. 2015.

\bibitem{Gupta_mm_MS}
S.~{Gupta}, T.~{Tomura}, S.~{Sakurai}, D.~J. {King}, and J.~{Hirokawa},
  ``Millimeter-wave {H}uygens’ metasurfaces based on all-dielectric
  resonators for antenna beam-forming,'' in \emph{2019 13th European Conference
  on Antennas and Propagation (EuCAP)}, Mar. 2019, pp. 1--3.

\bibitem{Kerker_Scattering}
M.~Kerker, \emph{The Scattering of Light and Other Electromagnetic
  Radiation}.\hskip 1em plus 0.5em minus 0.4em\relax Academic Press, New York,
  1969.

\bibitem{Howes_AD}
A.~Howes, W.~Wang, I.~Kravchenko, and J.~Valentine, ``Dynamic transmission
  control based on all-dielectric {Huygens} metasurfaces,'' \emph{Optica},
  vol.~5, no.~7, pp. 787--792, Jul 2018.

\bibitem{Nuaimi_AD}
M.~K.~T. {Al-Nuaimi}, Y.~{He}, and W.~{Hong}, ``Design of inhomogeneous
  all-dielectric electromagnetic-wave diffusive reflectarray metasurface,''
  \emph{{IEEE Antennas Wirel. Propagat. Lett.}}, vol.~18, no.~4, pp. 732--736,
  April 2019.

\bibitem{Decker_AD}
M.~Decker, I.~Staude, M.~Falkner, J.~Dominguez, D.~N. Neshev, I.~Brener,
  T.~Pertsch, and Y.~S. Kivshar, ``High-efficiency dielectric {Huygens}’
  surfaces,'' \emph{Adv. Opt. Mater.}, vol.~3, no.~6, pp. 813--820, 2015.

\bibitem{Saman_Huygen}
S.~Jahani and Z.~Jacob, ``All-dielectric metamaterials,'' \emph{Nat.
  Nanotechnol}, vol.~11, pp. 23--36, Jan. 2016.

\bibitem{Achouri}
K.~Achouri, A.~Yahyaoui, S.~Gupta, H.~Rmili, and C.~Caloz, ``Dielectric
  resonator metasurface for dispersion engineering,'' \emph{{IEEE Trans.
  Antennas Propagat.}}, vol.~65, no.~2, pp. 673--680, Feb. 2017.

\bibitem{Shengli_MS}
S.~Jia, X.~Wan, X.~Fu, Y.~Zhao, and T.~Cui, ``Low-reflection beam refractions
  by ultrathin {Huygens} metasurface,'' \emph{AIP Advances}, vol.~5, p. 067102,
  Jun. 2015.

\bibitem{Xue_Huygens}
C.~{Xue}, Q.~{Lou}, and Z.~N. {Chen}, ``Broadband double-layered {H}uygens’
  metasurface lens antenna for {5G} millimeter-wave systems,'' \emph{{IEEE
  Trans. Antennas Propagat.}}, pp. 1--1, 2019.

\bibitem{Kim_Huygens}
M.~Kim, A.~M.~H. Wong, and G.~V. Eleftheriades, ``Optical {H}uygens'
  metasurfaces with independent control of the magnitude and phase of the local
  reflection coefficients,'' \emph{Phys. Rev. X}, vol.~4, p. 041042, Dec. 2014.

\bibitem{GrbicmmWave}
C.~Pfeiffer and A.~Grbic, ``Millimeter-wave transmitarrays for wavefront and
  polarization control,'' \emph{Microwave Theory and Techniques, IEEE
  Transactions on}, vol.~61, pp. 4407--4417, 12 2013.

\bibitem{mmWaveHologram}
J.~Burch, J.~Ma, R.~I. Hunter, S.~A. Schulz, D.~A. Robertson, G.~M. Smith,
  J.~Wang, and A.~Di~Falco, ``Flexible patches for mm-wave holography,''
  \emph{Applied Physics Letters}, vol. 115, no.~2, p. 021104, 2019.

\bibitem{Sun}
Y.~Sun and K.~W. Leung, ``Millimeter-wave substrate-based dielectric
  reflectarray,'' \emph{{IEEE Antennas Wirel. Propagat. Lett.}}, vol.~17,
  no.~12, pp. 2329--2333, Dec. 2018.

\bibitem{PhaseshiftingSurface}
C.~{Pfeiffer} and A.~{Grbic}, ``Millimeter-wave transmitarrays for wavefront
  and polarization control,'' \emph{{IEEE Trans. Microw. Theory Tech.}},
  vol.~61, no.~12, pp. 4407--4417, 2013.

\bibitem{Emara}
M.~Emara, T.~Tomura, J.~Hirokawa, and S.~Gupta, ``Reflection-cancelling
  dielectric {H}uygens' metasurface pair for wideband millimeter-wave
  beam-forming,'' \emph{2019 IEEE Antennas and Propagation Society
  International Symposium}, Jul. 2019.

\bibitem{SingleDielectricHUygen}
H.~{Lv}, Q.~{Huang}, X.~{Yi}, J.~{Hou}, and X.~{Shi}, ``Low-profile
  transmitting metasurface using single dielectric substrate for oam
  generation,'' \emph{{IEEE Antennas Wirel. Propagat. Lett.}}, pp. 1--1, 2020.

\bibitem{MS_rev_Transmit}
S.~Bukhari, J.~Vardaxoglou, and W.~A. Whittow, ``Metasurfaces review:
  Definitions and applications,'' \emph{Appl. Sci,}, vol.~9, p. 2727, 2019.

\bibitem{Sherman_Monopulse}
S.~Sherman and D.~Barton, \emph{Monopulse Principles and Techniques}, ser.
  Artech House radar library.\hskip 1em plus 0.5em minus 0.4em\relax Artech
  House, 2011.

\end{thebibliography}

\end{document}